\documentclass{article}
\usepackage{graphicx} %
\usepackage[font=footnotesize]{caption}
\usepackage[utf8]{inputenc}
\usepackage{authblk}
\usepackage{booktabs}
\usepackage{verbatim}
\usepackage{url}
\usepackage{caption}
\usepackage{subcaption}
\usepackage{indentfirst}
\usepackage{url}
\usepackage{amssymb}
\usepackage{amsmath}
\usepackage{amssymb}
\usepackage{bbold} 
\usepackage{xcolor}
\usepackage{hyperref}

\title{Some Considerations on Measuring and Comparing Things}

\author[1]{Luciano da F. Costa$^*$}
\author[1]{Gonzalo Travieso}
\author[2]{Alexandre Benatti}

\affil[ ]{\hspace*{-.15cm}\raggedright $^*$ Prof. Senior. \newline}

\affil[1]{S\~ao Carlos Institute of Physics - DFCM, University of S\~ao Paulo, Av.~Trabalhador S\~ao-carlense, 400, S\~ao Carlos, SP 13566-590, Brazil \vspace{.5 cm}}

\affil[2]{Institute of Mathematics and Statistics - DCC, University of S\~ao Paulo, Rua do Mat\~ao, 1010, S\~ao Paulo, SP 05508-090, Brazil}

\date{22nd Sep. 2026}

\begin{document}

\maketitle

\begin{abstract}
Measurements and relationships between measurements are basic aspects underlying science and technology. In the present text, we present and discuss some of the approaches for measuring,  normalizing, and comparing entities represented in terms of a set of measurements.  Three main categories of measured values (constant, direct, or inverse) are proposed, underlaid by the relationship between data variation or intervals of interest and magnitude of the quantifications. For example, in certain cases the variation of measurements can be expected to be proportional to the relative magnitude (e.g.~when the magnitudes extend over long ranges). The considered approaches to comparing measurements include the Euclidean distance, three types of proximity based on the latter, as well as the Jaccard and coincidence indices. The presentation is complemented by graphical illustrations.  
\end{abstract}

\section{Introduction}

Science and Technology are founded on the quantification of physical properties of entities or dynamics (e.g.~\cite{baird1995,cataldo2020,morris2020,holman2012}), which contribute not only to enhanced objectivity of the developed approaches but also to their respective continued validation. In particular, it is by measuring properties of a given object that it can be represented and modeled in mathematical terms. These quantifications also pave the way to relating natural entities and dynamics which, as a matter of fact, is an operation often employed in scientific modeling.

Inferring relationships between objects in nature involves the adoption of some means for estimating distances and similarities between measurements of their respective properties. As an example, weather prediction involves comparing several meteorological measurements over time, places, and databases. However, the estimation of relationships in science and nature goes much further and involves virtually every aspect of related concepts and methodologies. In particular, the validation of scientific models relies on \emph{comparing} and \emph{relating} simulation results (predictions) with quantities measured from real-world experiments.

\begin{figure}[ht]
  \centering
     \includegraphics[width=0.99 \textwidth]{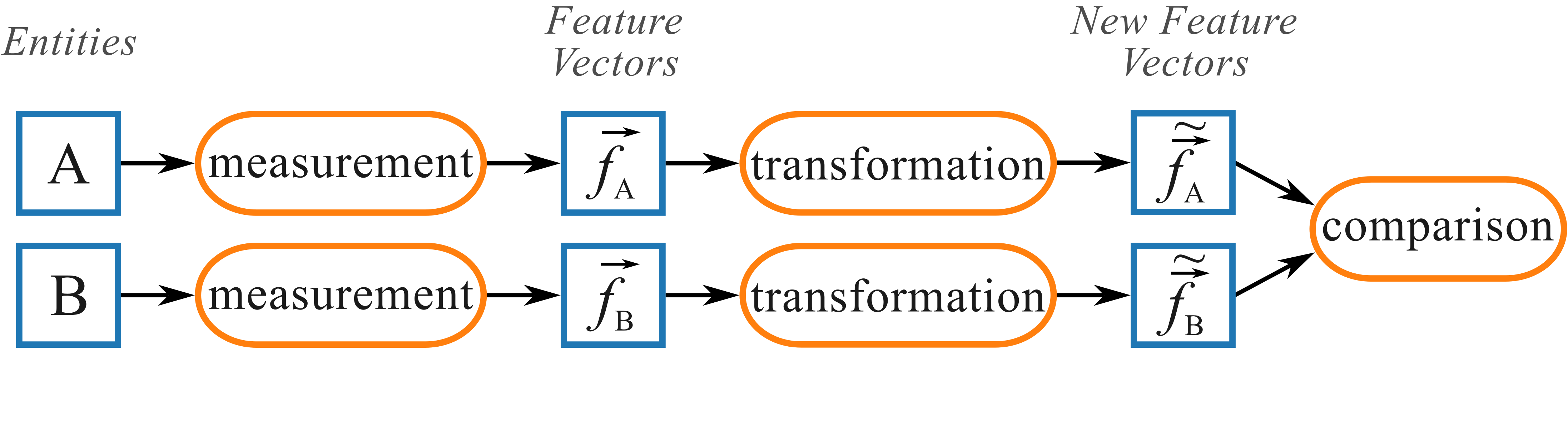} 
  \caption{Diagram illustrating the measurement of two entities $A$ and $B$, yielding respective feature vectors which can then be transformed into new feature vectors. Frequently in science and technology, the obtained feature vectors need to be compared.  Interestingly, measurements can be understood as transformations while measurements often correspond to comparisons with a reference, or standard.} \label{fig:diagram}
\end{figure}

Figure~\ref{fig:diagram} illustrates the above concepts respectively to the characterization and comparison of two entities $A$ and $B$, each of  which represented in terms of a respective feature vector (measurements).  After eventual transformations, the new feature vectors are compared by using some suitable approach.

The present work is intended to provide an introduction to some of the main approaches which can be used to quantify the relationship between properties represented in terms of a given property $x$, which is also called a feature, attribute, characteristic, or measurement.  For simplicity's sake, only a single feature (or measurement) is considered henceforth.

The described development organizes measured values into three main groups according to their implied relationship between quantity variation, dispersions (or intervals of interest) $\delta > 0$ and the magnitude $|x|$ of the property $x$. The quantity $\delta$ is henceforth employed to indicate a property that is expected for each magnitude of the obtained measurements.  As such, it can have different meanings, which need to address each specific situation.  For instance, there are situations, as when measuring objects with markedly different sizes (resolving power), in which it is interesting to take $\delta$ as being proportional to the magnitude.  However, the value $\delta$ could also relate to how the length of intervals of interest scale with the magnitude, or in situations in which the dispersion of interferences (e.g.~noise) is known to be proportional to the measured values (e.g.~Poisson noise, e.g.~\cite{snyder1991}), as is often the case in electronics and some types of imaging.

The three types of measurement data are here called \emph{uniform}, \emph{direct}, and \emph{inverse}. The first is characterized by $\delta$ being constant, while in the second case it increases with the magnitude. When this increase is linear, in the current work, the measured values are said to be \emph{proportional}~\cite{benatti2024}. The third type of measured value, characterized by $\delta$ expected to vary inversely with the magnitude, is less common and can often be addressed in terms of proportional measurements; they will not be covered in more detail here.  The important subject of measurement normalization is also addressed in the present work with respect to uniform and proportional types of measurements.

Once properties of entities have been measured, they often need to be related, which can be done in terms of \emph{comparisons}. In the present work, three main approaches to comparing measurements are covered: (i)  \emph{distance} (Euclidean); (ii)  \emph{proximity} (three possible approaches); and (iii) \emph{similarity} (Jaccard and coincidence indices).  In this work, the latter index is considered only respectively to the comparison between two positive values.

The Euclidean distance is probably the most frequently considered means for comparing and relating data due to its intrinsic compatibility with human intuition: we live in a 3D world where distances are particularly important. The Euclidean distance will be the first approach addressed in this work, also providing a kind of reference against which other approaches can be related. Although there is a virtually infinite number of possibilities for defining distances, here only the Euclidean distance is addressed.

Instead of distances, it is also possible to consider the \emph{proximity} between data elements. Perhaps one of the most direct approaches to proximity is to consider some way in which the Euclidean distance is transformed into a measurement of proximity. These possibilities include taking the complement of the distance, its reciprocal value, or taking the exponential to a negative distance exponent. These three possibilities are considered and illustrated here.  

Another interesting family of approaches focuses on the concept of \emph{similarity}. Among the most frequently applied similarity measurements we have the magnitude of the \emph{inner product} between two feature vectors.  More specifically, two vectors that are aligned tend to produce larger values of this magnitudes, which is also proportional to the product of the vectors magnitude.  Although the \emph{Jaccard similarity index} has been frequently used for comparing sets, it can be generalized for the comparison of two vectors.  In this case, it remains normalized in the interval $[0,1]$, the Jaccard index (e.g.~\cite{Jac:wiki,da2021further}), being particularly suitable to compare proportional measurements~\cite{benatti2024,benatti2024partially}. The quantification of similarity, which can also be obtained in terms of the coincidence indices, is also presented and illustrated.

This work begins by reviewing and discussing preliminary concepts including units, bounding, dimensionality, and statistical models (distributions). Next, three types of measured values are discussed with respect to how their local dispersion tends to vary with magnitude. The work follows by addressing the Euclidean distance, three types of proximity, and the Jaccard and coincidence similarity indices.

\section{Preliminary Considerations}

Before measuring and relating entities, several aspects need to be carefully taken into account. This section briefly reviews some of them.

First, it is interesting to observe that the word \emph{measurement} can mean the mathematical aspects of quantification (e.g.~linear, logarithmic), the act of quantifying a property by using a given instrument (e.g.~\cite{baird1995,cataldo2020,morris2020,holman2012}), or the measured value itself (e.g.~$10kg$), which is the quantification of the property of the entity of interest. In the present work, the term \emph{measured value} has been preferably used to indicate the latter meaning.

Another aspect to be kept in mind is that, in practice, each type of measurement is \emph{relative} to an original space (or previous measured value), which, frequently, is linear.  For example, a logarithmic scale of measurement $\log(x)$ assumes that $x$ is originally on a linear scale (e.g.~sizes of objects in a linear vector space). Actually, most measurements can be understood as \emph{transformations} of the respective original quantities. Henceforth, unless otherwise noted, the properties to be quantified are assumed to exist in a linear space, while the implemented measurements are understood to correspond to transformations obtained by using a given means (e.g.~instrument, counting, etc.). Interestingly, there are several situations in which the type of space in which the property is being quantified is not known, so that the only information is that obtained by applying the transformation implemented by the measurement. Once a quantity is measured, it can be understood as the property of the respective data element.

Some of the most important properties of \emph{measured values}  include: (a) their units (e.g.~\cite{Chertov1964}); (b) being bound or unbound; (c) their dimensionality; and (d) their statistical distribution. The present work focuses on measurements on one-dimensional data.

Among the simplest types of measurement are those obtained from linear vector spaces, such as positions $x$ along the continuous real line $\mathcal{R}$ measured by a ruler.

It should be kept in mind that there is an infinite number of ways in which the divisions of a scale can be implemented, and most of them can be understood as transformations of the uniform case:
\begin{align}
  \tilde{x} = f(x)
\end{align}
where $f(\cdot)$ is a generic function.

The choice of the transformation function $f(\cdot)$ should take into account several aspects. First, we have compatibility with the property and data to be studied; for instance, to use a logarithmic scale the original values must be positive. Then the measurement instrument needs to be taken into account, as it may provide measurements involving a composite transformation function. For example, the Richter and decibel scales are logarithmic.

Consider a given data set $A$ characterized by a single feature uniformly measured as $x$. If, after measuring several samples, it is found that the data elements tend to present an exponential distribution along $x$, the transformation by the inverse function $f(x) = log(x)$ could be considered as a means to obtain a more uniform distribution of observations.

Generally speaking, the choice of $f(x)$ is an important first step to consider in light of several aspects, including the nature of the data, the measurement devices, and the main questions and constraints underlying each study.

Once a measurement is specified, it is important to take into account its unit. For example, measurements of distances could have units of centimeters or inches, among many other possibilities. At the same time, measurements of temperature could be in terms of degrees Fahrenheit or Celsius. Unit changes can also be understood as measurement transformations. There are measurements which are dimensionless, such as cardinality (numbers of elements). Another important type of dimensionless measurement $x$ is relative to a reference $x_r$:
\begin{align}
  \hat{x} = \frac{x}{x_r}
\end{align}

As an example, we could be interested in measuring the dispersion of a measurement, quantified by its standard deviation, with respect to its average:
\begin{align}
  \hat{x} = \frac{\sigma_x}{\mu_x}
\end{align}
This measurement, in particular, is called \emph{coefficient of variation}.

Another important property of a given measurement concerns whether it is \emph{bound} or \emph{unbound}. As an example, the measurement of distances, from a mathematical point of view, unbound. At the same time, the measurement of the volume of a liquid in a bottle is bound by the total capacity of the bottle.

\section{Three Types of Measured Values }

In practice, the choice of measurements is often influenced by existing instruments capable of performing a given measurement, unless a new type of instrument needs to be implemented. This aspect is related to traditionally established measurements (e.g.~sound intensity is typically measured in decibels).

Another important factor to be taken into account concerns the \emph{expected variation} (or \emph{dispersion}, or interval of interest) of the measurement as a function of its magnitude. In general, the dispersion of measurements of the sizes of a type of real world object can be expected to be proportional to their average magnitude. As an example, the variation of the size of watermelons can be expected to be larger than that obtained while measuring peanuts.  Relative measurements (e.g.~percentage) are typically adopted in these situations, so that the same $10\%$ variation of the measurement of the size of a watermelon or peanut corresponds to markedly distinct absolute variations. This interdependence between measurement dispersion and magnitude ultimately influences many applications of measurements.

The resolution in which a measurement at a given magnitude can be detected or displayed could also be related to the density of objects to be measured around a reference magnitude. In other words, the resolution could adapt so as to follow the density. For example, in case where a measured property is known to be normally distributed, with a peak at a given magnitude, the resolution of the respective measurement could be higher around that same magnitude, because the higher density may reflect greater proximity between the data elements at that point. It is also possible that, even in such situations, high resolution could be needed also for less dense intervals that deserve special attention.  

It should also be taken into account that the resolution of a measurement is not the same as its \emph{accuracy}, though these aspects are interrelated. Although lower resolution usually implies smaller accuracy, this aspect can also be influenced by other factors and considerations. Accuracy is a quantification of how close the measured property is to the measurement. Measurement errors are typically a consequence of visual scale misreading, noise, and other interferences (systematic or not). In this way, accuracy is more closely related to the means (instrument) employed for the measurement than to the choice of a particular type of measurement.

Another aspect related to the choice of type of measurement (and respective instrument) concerns \emph{sensitivity}, which indicates the differential variation of the measured value for a differential increase of the measured quantity. Typically, the sensitivity of a instrument could be expected to be higher in critical regions of the scale or specific intervals that require special attention. Other relationships between density and resolution are possible according to each specific situation.

Yet another aspect which can strongly influence the choice of a type of measurement concerns the way in which a given data set is supposed to be studied. For instance, in some cases a given type of measurement (e.g.~linear or logarithmic) can be preferred to be used regardless the data type and statistical properties. A possible example is the measurement of fruit sizes linearly by using a linear caliper irrespectively to density, because this approach is not only more traditional, but also allows for a more intuitive interpretation and discussion of the obtained results.

In the present work, measured values are organized into three main categories from the perspective of the expected dispersion or range of interest $\delta$ of the measurement $x$ in terms of its magnitude $|x|$. From this perspective, measured values can be organized into the following three main categories: (I) $\delta$ remains constant with magnitude; (II) $\delta$ increases directly with magnitude; and (III) $\delta$ increases reversely with magnitude. These categories of measured values are henceforth referred to as \emph{uniform}, \emph{direct}, and \emph{inverse}.

Probably the simplest among these three types is category I, which is characterized by measurement values in a linear space being uniformly distributed. Interestingly, examples of natural properties following this type of distribution are rare, and include the angle of molecular collisions and the orientation of galaxies in deep space. Although most natural properties are not uniformly distributed, they have traditionally been quantified by uniform types of measurement (category I).

Category II of measured values is more commonly found in science and technology. A particular example is measuring the size of fruits over a wide range of scales (e.g.~from berries to watermelons). Typically in these cases, the smaller the fruit, the smaller the expected variation. Another example is the measurement of the intensity of sound. As will be discussed in the present work, logarithmic measurements can be employed in these situations.

The use of measurements of category III are very rare, being mainly related to situations slowly reaching a limit value, as is the case with the relative purity of a given material, liquid, etc. These cases typically involve two components, a \emph{majoritarian} one (e.g.~the quantity of Silicon in semiconductors), and a \emph{minority} component (e.g.~the impurities in semiconductors). The properties of these compounds are known to vary abruptly with small increases of impurity level (e.g.~a semiconductors start to conduct significantly with the addition of small fractions of impurities). In the present work only uniform and proportional measurements are further addressed and discussed.

\begin{figure}[ht]
  \centering
     \includegraphics[width=0.9 \textwidth]{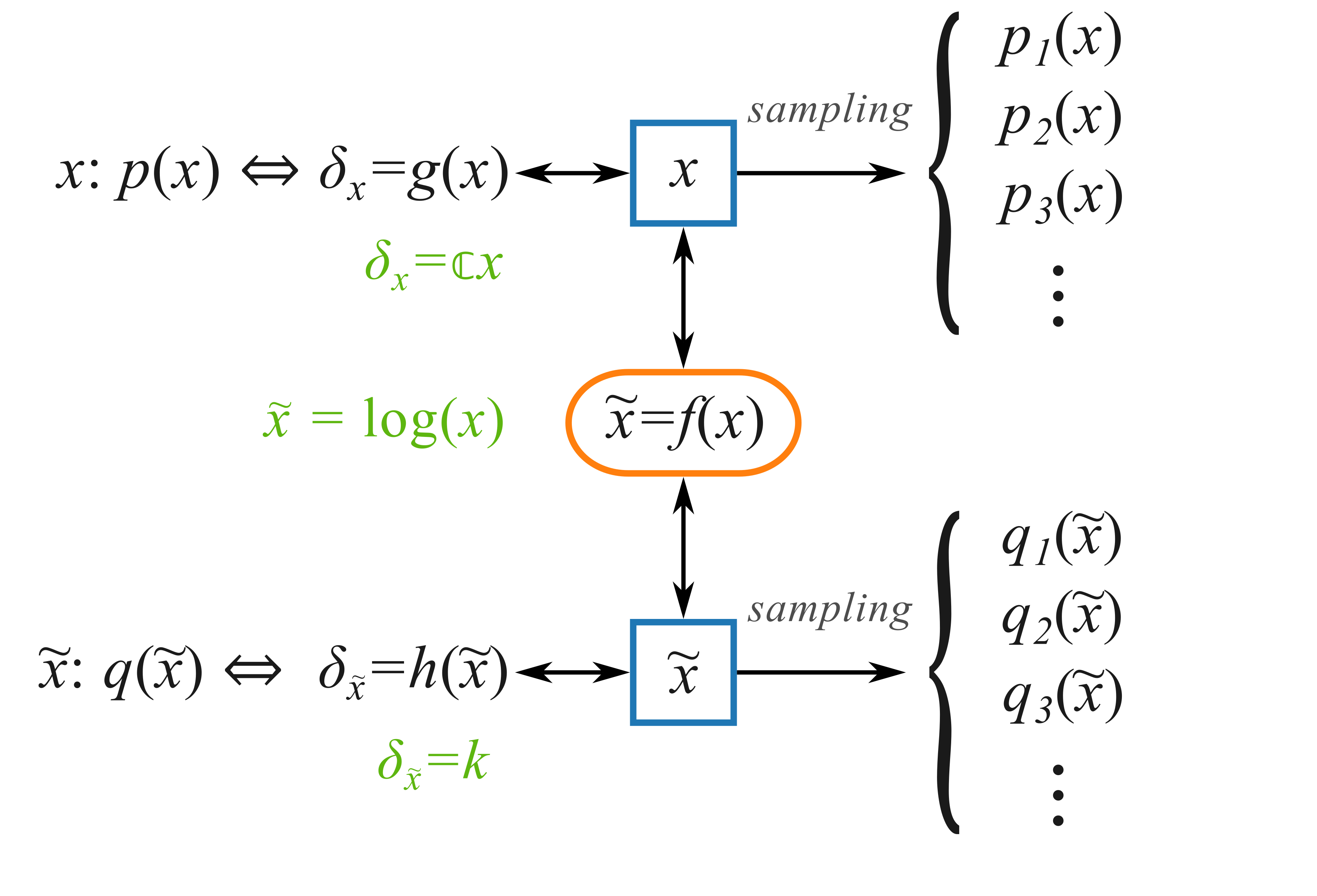} 
  \caption{Diagram indicating some of the many factors related to a given measurement, which is here understood as the transformation $f(\cdot)$ of the original quantity $x$ in the measured values $\tilde{x}$. Observe that $\mathbb{c}$ and $k$ are constant values.} \label{fig:diagram2}
\end{figure}

Figure~\ref{fig:diagram2} illustrates some of the several aspects involved in measurements, with emphasis on the effect of the expected variation, dispersion or interval of interest $\delta$.

The original data has statistical density $p(x)$, which is associated to a statistical model involving the density $g(x)$.  More specifically, the relationship between $\delta$ and $|x|$ can be enough to specify the density of probability modeling the random variable $x$. In practice, it is only possible to estimate $p(x)$ in terms of approximated densities $p_1(x), p_2(x), \ldots$ obtained by using different data samples or estimation approaches. These estimations will depend strongly on the number of samples and the ranges of interest along the $x$ axis. The measurement $\tilde{x}=f(x)$ implies a new relationship $\delta_{\tilde{x}} = h(\tilde{x})$, which is associated to a respective model density $\tilde{x}: q(\tilde{x})$, which can be sampled as $q_1(\tilde{x}), q_2(\tilde{x}), \ldots$. In case the transformation $f(x)$ is known, these approximated densities can be obtained mathematically from respective densities $p_1(x), p_2(x), \ldots$.

It is interesting to observe that $g(\cdot)$ and $h(\cdot)$ cannot necessarily be obtained by sampling the respective measured values unless a large number of samples is obtained considering the whole range of the observe quantities.  Also shown in Figure~\ref{fig:diagram2} (in green) is a particular example in which $\delta_x = \mathbb{c} x$, which can be implemented in terms of the logarithmic function.

The three categories described above are henceforth considered in the present work, providing a rationale for the characterization of not only the discussed measurements, but also of their transformations, normalizations, and comparisons.

Though the measurement values in the three types of categories can be modeled in a virtually infinite number of ways, the following reference mathematical models are taken into account henceforth.
\begin{align}
    &\delta = \mathbb{c} \\
    &\delta = \mathbb{c} \, |x| \label{eq:prop_meas} \\  
    &\delta = \frac{\mathbb{c}}{|x|}
\end{align}

A direct measurement as in Equation~\ref{eq:prop_meas} above is henceforth called a \emph{proportional} measurement.

\begin{figure}[ht]
  \centering
     \includegraphics[width=0.99 \textwidth]{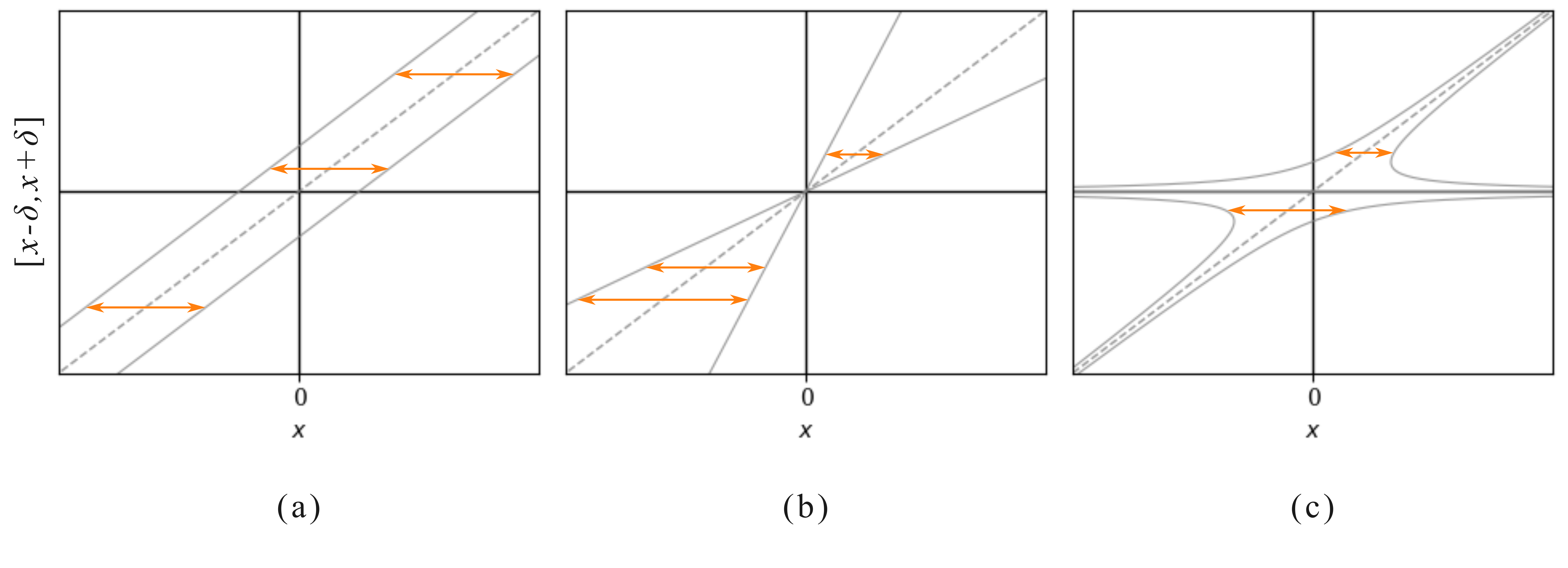} 
  \caption{Graphic illustration of the variation intervals (in orange) implied by the uniform (a), proportional (b), and inverse (c) measured values. To avoid overlap, the intervals are vertically translated by the identity function.} \label{fig:ratio}
\end{figure}

Figure~\ref{fig:ratio} illustrates the measurement value variations characterizing each of the three categories considered. This graph indicates, along the horizontal direction, the ranges of variation of $x$ for each value of $x$ transported by the identity function to the vertical axis. In the case of category I, shown in (a), the dispersion or interval of interest $\delta$ remains constant whatever its magnitude $|x|$. Data measurement values in category II are characterized by $\delta$ increasing (not necessarily in a linear manner as in the figure) with the magnitude. In case of measured values of type III, the dispersion or interval of interest decreases with the magnitude.  

\section{Uniform Measured Values}

As discussed previously, uniform measured values are characterized by constant local dispersion or interval of interest. They are particularly associated to properties that are linearly distributed, but have been used in other types of data as well.   In particular, this type of transformation does not change the mathematical form of the distribution of the original quantity, although it typically changes the magnitude of the measured data.

Given a set $X$ of measured values $x$, properties of special interest include:
\begin{align}
    &x_{min} = \min(X)  \\
    &x_{max} = \max(X)  \\
    &\mu_X \\
    &\sigma_X
\end{align}
where $\mu$ and $\sigma$ are the average and standard deviation of the values $x$.

It frequently happens that uniform measured values need to be normalized in order to become invariant to translation or scaling of the coordinate system.  For instance, the normalization:
\begin{align}\label{eq:norm1}
    \tilde{x_i} = x_i - x_{min}
\end{align}
standardized the minimum value of the transformed measurement to 0.

A frequently employed normalization is in terms of the linear ratio:
\begin{align}\label{eq:norm2}
   \tilde{x} = \frac{x- x_{mim}}{x_{max}-x_{min}}
\end{align}
where $x_{min}$ and $x_{max}$ are, respectively, the minimum and maximum value of $x$, both of which are assumed to be finite. It can be verified that $\tilde{x}$ is dimensionless and bounded in the interval $[0,1]$.

Because normalizations can be understood as a particular type of measurement transformation, every observation made about measurement transformations in the previous section applies directly to normalization. For example, there are also an infinite number of ways in which a measurement can be normalized.  

Another interesting normalization of statistical nature that is often considered is as follows:
\begin{align}\label{eq:norm3}
   \hat{x} = \frac{x - \mu_x}{\sigma_x}.
\end{align}
Although the bounds of the new dimensionless measurement $\hat{x}$ are not fixed,  most observations of $\hat{x}$ result between $-2$ and $2$.

As with measurement transformations, the choice of normalizations should also be performed carefully while considering several aspects of the data and the intended analysis.

\section{Proportional Measured Values and the Log-Uniform Distribution}

Direct measurements have $\delta$ increasing with the magnitude of the quantities $x$ being measured. Though this increase can occur in several ways, in the present work we focus on proportional measured values, i.e.~$\delta = c \, |x|$. In this case, the measurements are typically implemented in terms of a linear scale.  

A way to normalize proportional measurements (e.g.~\cite{benatti2024}), considering that all values $x_i$ are non-negative, is as follows:
\begin{align}
    \hat{x} = \frac{x}{\rho}
\end{align}
where:
\begin{align}
    \rho = \sqrt{\frac{1}{N}\sum_{i=1}^N x_i^2}
\end{align}

The normalizing factor $\rho$ can be understood as a measurement of the dispersion of the values $x_i$ respectively to the origin of the coordinate system. As a consequence, the normalized values $\hat{x}$ will have this type of dispersion equal to $1$. Because $\rho$ has the same unit as $x$, the normalized values $\hat{x}$ become dimensionless.

The proportionality of measured values is directly related to the \emph{log-uniform} distribution (also called \emph{reciprocal} distribution), which is described as follows.

Let us start with a uniform distribution specified as:
\begin{equation}
p(x) = 
\begin{cases} 
\frac{1}{d-c}, & \text{if } c \leq x \leq d \\ 
0, & \text{otherwise} 
\end{cases}
\end{equation}
with $c > 0$.

A new random variable $Y$ can now be obtained by transforming $X$ as follows:
\begin{align}
    y = e^x, \; \; a = e^c, \; b = e^d
\end{align}
which has probability density function as follows:
\begin{align}
    p(y) = \frac{1}{y \, \ln(b/a)}, \; \text{for } a \leq y\leq b
\end{align}

The respective cumulative distribution function can be expressed as:
\begin{align}
    P(y) = \frac{\ln(y/a)}{\ln(b/a)}
\end{align}

\begin{figure}[ht]
  \centering
     \includegraphics[width=0.5 \textwidth]{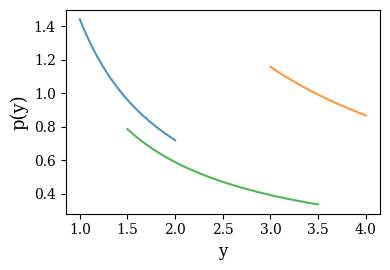}
  \caption{Examples of log-uniform densities for $a=1$ and $b=2$; $a=3$ and $b=4$; and
$a=2$ and $b=4$.}
  \label{fig:logUniform}
\end{figure}

Figure~\ref{fig:logUniform} illustrates log-uniform densities obtained for $a=1$ and $b=2$; $a=3$ and $b=4$; and
$a=2$ and $b=4$.

The log-uniform distribution has especially interesting properties, including: (a) the probability of drawing a value in the intervals $[1,10]$; $[10,100]$; and $[100,1000]$ (among infinite other possibilities) is precisely the same; (b) if the values $a$ and $b$ are scaled by a positive constant $k$, the shape of the density function does not change.

Now, if we start with $Y$ and make the inverse transformation:
\begin{align}
    x = \ln(y),   \; \text{for }  c = \ln(a), \; d = \ln(b) 
\end{align}
the resulting variable $X$ is characterized by a uniform probability density function.

This result illustrates the possible use of the logarithmic transformation as a means of uniformizing a measurement.

\section{On the Invariant Properties of Comparisons}

So far in this text, we have presented and discussed two main types of measured values, namely uniform and proportional.  The choice of approach to compare these types of values should take into account several aspects not only from the own data but also of what is expected of the subsequent comparisons, analysis, and modeling.  

One particular aspect influencing the choice of a comparison approach concerns how the comparison acts on the relationship between $\delta$ and the magnitude of $x$.  When the two values $x_1$ and $x_2$ to be compared are of category I, it is typically expected that the comparisons $c(x_1,x_2)$ be invariant to joint translation by $a\in \mathcal{R}$ of the values, i.e.:
\begin{align}
  c(x_1,x_2) = c(x_1 + a, x_2 + a)
\end{align}

It can be verified that the Euclidean distance (see Section~\ref{euclidian}) satisfies this requirement:
\begin{align}
  |x_2 - x_1| = |(x_2 + a) - (x_1 + a)|    
\end{align}

Another property expected from comparisons between uniform measurement values is that it would scale linearly with the scaling of the values $x_1$ and $x_2$, a property that is also satisfied by the Euclidean distance.  Taking $s\in \mathcal{R}$ as the scaling factor, we would have:
\begin{align}
    c(s x_1, s x_2) = s\,c(x_1, x_2).
\end{align}
This is also satisfied by the Euclideian distance:
\begin{align}
  |s\, x_2 - s \, x_1| = s |x_2-x_1|.
\end{align}

Therefore, the properties of the Euclidean distance, allied with its intuitive interpretation, make it particularly suitable for comparing uniform measured values, although other comparison approaches can also be considered for each specific situation.

In the case of proportional measured values, the comparison between two values $x_1$ and $x_2$ is expected to be invariant to their joint scaling by $s$:
\begin{align}
    c(s x_1, s x_2) = c(x_1, x_2).\label{eq:propcomp}
\end{align}

The Jaccard similarity index (see Section~\ref{jaccard}) can be verified to satisfy this requirement:
\begin{align}
    \mathcal{J}(s\, x_1,s \, x_2) = \frac{\min(s \, x_1, s\, x_2)}{\max(s\, x_1,s\, x_2)}
    = \frac{s\, \min(x_1,x_2)}{s\, \max(x_1,x_2)} =
     \frac{\min(x_1,x_2)}{\max(x_1,x_2)} = \mathcal{J}(x_1,x_2)
\end{align}

It should be observed that the Jaccard similarity is not invariant to translation.

Another interesting property of the Jaccard similarity is that taking both compared values to the $d-$th power (with $d>0$) will imply the similarity to be taken to that same power, i.e.:
\begin{align}
    \mathcal{J}(x_1^d, x_2^d) = \frac{\min(x_1,x_2)^d}{\max(x_1,x_2)^d} =
     \left[\frac{\min(x_1,x_2)}{\max(x_1,x_2)}\right]^d = \left[ \mathcal{J}(x_1,x_2) \right]^d
\end{align}

The properties of the Jaccard similarity index make it a potentially interesting choice for comparing proportional measured values. 

The Euclidean distance (in one dimension) as well as the Jaccard similarity index are discussed in more detail in the following sections of this text.

\section{The Euclidean Distance}\label{euclidian}

The comparison between linear measured values is frequently implemented in terms of the \emph{Euclidean distance}, which is defined for geometric linear vector spaces of any dimension as:
\begin{align}
    E(\vec{v}, \vec{u}) = \sqrt{ \sum_{i=1}^{N} \left(x_{i,v}-x_{i,u} \right)^2 +
    \left(y_{i,v}-y_{i,u} \right)^2}
\end{align}

In this work, we will focus on the one-dimensional form of the Euclidean distance between two values $x_1$ and $x_2$, which is:
\begin{align}
    E(x_1, x_2) = | x_1 - x_2 |
\end{align}

It can be verified that this quantity obeys the three properties required for a formal definition of distance, namely being non-negative, commutative, and satisfying the triangular inequality.

The Euclidean distance is, in principle, unbounded.  However, it can be considered bounded by the largest measurement that can be observed in a given experiment. In these cases, it is possible to consider normalization schemes aimed at standardizing the range of data variation.  Normalization schemes more directly related to the Euclidean distance are as in Equations~\ref{eq:norm1}, ~\ref{eq:norm2}, and~\ref{eq:norm3}.

\section{Proximity}

In this work, proximity $P(x_1,x_2)$ is understood here as a measurement that quantifies how close two values $x_1$ and $x_2$ are. The considered approaches are expressed in terms of transformations of the  Euclidean distance. This type of comparison is motivated as a means of understanding proximity in terms of the intuitive and well-known concept of Euclidean distance, which is particularly familiar to humans.

The simplest way to transform the Euclidean distance into a proximity measurement consists of taking its complement with respect to a reference value $M \geq \max(E(x_1, x_2))$.  In this case, we have:
\begin{align}
    P_1(x_1, x_2) =  1- \frac{E(x_1, x_2)}{M}
\end{align}

The choice of $M$ depends on each specific situation and the range of interest $x_1, x_2 \in [x_{min},x_{max}]$, and imposes a constraint to choose $M$, namely $M \geq x_{max}-x_{min}$. The value of $M$ also specifies how strict is the proximity comparison. More specifically, stricter comparisons are obtained for smaller values of $M$. In addition, observe that $P_1$ is dimensionless.

This type of proximity is bilinear, implying no warping to the original space. 

One approach which is frequently adopted in the literature consists of taking the inverse of the Euclidean distance, i.e.:
\begin{align}
    P_2(x_1, x_2) =  \frac{1}{E(x_1,x_2)}
\end{align}

In this case, the proximity remains unbound, while the original unit becomes reciprocal, and non-uniform changes are implied onto the scale.

This type of proximity is not linear, presenting a singularity that implies the proximity to increase to infinite as the distance goes to zero.  At the same time, the proximity converges to zero for infinite distances. Therefore, this measurement applied to an unbound set of values will also be unbound.

A third alternative for transforming the Euclidean distance in a proximity measurement is as follows:
\begin{align}
    P_3(x_1, x_2) =  e^{-\alpha E(x_1,x_2)}
\end{align}

This measurement has the interesting property of being bound in the interval $[0,1]$. However, strictly speaking, it is not suitable to take the exponential of a non-dimensionless quantity. A possible way to bypass this difficulty is to consider the constant $\alpha$ as having unit that is reciprocal to the distance unit.

The proximity $P_3$ acts similarly to $P_2$, but the intensification of the comparison is attenuated when the compared values are small.

\begin{figure}[ht]
  \centering
     \includegraphics[width=0.99 \textwidth]{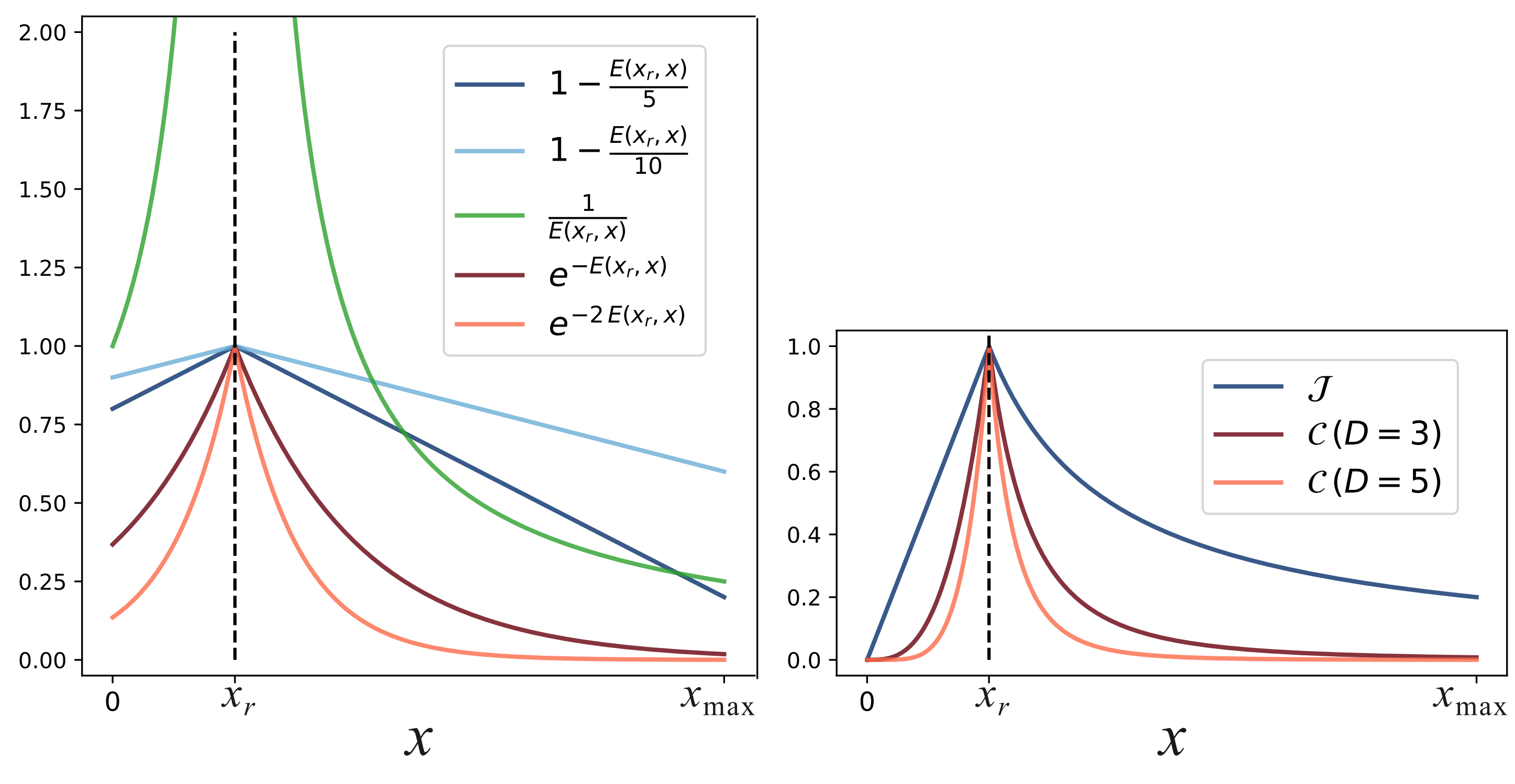}\\
     \hspace{.6 cm} (a) \hspace{5.5 cm} (b)
  \caption{Comparison of alternative proximity and similarity between $x$ and $x_r=1$. (a) Proximity values obtained by using the first type of proximity (with $M=5, 10$), the second type, and the third type (with $E_R=1$ and $\alpha = 1, 2$). (b) Values obtained by using the Jaccard similarity and the coincidence similarity (with $D=3, 5$).}
  \label{fig:prox_sim}
\end{figure}

Figure~\ref{fig:prox_sim}(a) presents a comparison between the three types of proximity considered in this work with respect to a reference point $x_r=1$, in the range $x_{min}=0$ and $x_{max}=5$.  Two values of $M$ ($M= 5$ and $M= 10$) have been taken for the first type of proximity, and two values of $\alpha$ have been considered for the third type of proximity.

One of the most important points to be observed is that completely different values of proximity can be obtained by using each of the considered measurements and respective parameter configurations. Therefore, special care must be taken when choosing a proximity approach, considering several aspects, including the type of data and the questions being addressed, among other factors. At the same time, the obtained curves are mutually congruent in the sense of presenting the same \emph{monotonicity}, therefore tending to lead to similar \emph{qualitative} analysis.

The curves in Figure~\ref{fig:prox_sim}(a) also emphasize the different types of axis warping implemented by each of the types of proximity, $P_1$ being characterized by uniform (linear) variation.

\section{The Jaccard and Coincidence Similarity Indices}\label{jaccard}

Given two sets $A$ and $B$, the Jaccard similarity index between them is defined as:
\begin{align}
    J(A,B) = \frac{|A \cap B|}{|A \cup B|}
\end{align}
where $|\cdot|$ means the cardinality of a set, corresponding to the respective number of elements.

Because the maximum value of the cardinality of the intersection of A and B is limited to the cardinality of the respective union, the Jaccard index is bound in the interval $[0,1]$, which is particularly convenient in many applications.

The Jaccard index can be extended to vector spaces by considering a multiset (e.g.~\cite{Blizard,Blizard2,da2021further}) approach in which the union and intersection of two vectors are implemented in terms of the maximum and minimum operations.  More specifically, in a one-dimensional space of \emph{non-negative values} we have:
\begin{align}
    \mathcal{J}(x,y) = \frac{\min(x,y)}{\max(x,y)}
\end{align}
which is also bounded in the interval $[0,1]$.

The Jaccard similarity between proportional measurements $x$ does not depend on the magnitudes $x_1$ and $x_2$, but only of their proportional ratio $\mathbb{c}$ (or separation $\delta = c \, |x|$). This can be illustrated by considering its application to compare pairs of values $(x_1, x_1+ \mathbb{c} \,x_1)$ and $(x_2,  x_2+ \mathbb{c} \, x_2)$; for $x_1 \neq x_2$, $x_1,x_2 \geq0$; and $\mathbb{c}>0$. Observe that in this case, both pairs of values are separated by dispersions which are proportional to the magnitude of the smallest value, i.e.~ $\delta_{x_1} = \mathbb{c} x_1$ and  $\delta_{x_2} = \mathbb{c} x_2$. By applying the above equation, we have:
\begin{align}
    &\mathcal{J}(x_1, x_1+ \mathbb{c} \,x_1) = \frac{x_1}{x_1+ \mathbb{c} \, x_1 } =  \frac{1}{1+ \mathbb{c} } = \kappa  \\
    &\mathcal{J}(x_2,  x_2+ \mathbb{c} \, x_2) = \frac{x_2}{x_2+ \mathbb{c} \, x_2 } = \frac{1}{1+ \mathbb{c} } = \kappa   
\end{align}
i.e.~the Jaccard similarity comparison between proportional measurements with the same $\mathbb{c}$ is constant and equal to $\kappa = 1/(1+\mathbb{c})$.  Therefore, more separated pairs of proportional values (i.e.~$\mathbb{c}$ large) produce a smaller Jaccard index.  At the same time, comparisons between more proximate proportional values (i.e.~$\mathbb{c}$ smaller) will result in larger comparison values, which is expected since the Jaccard index implements similarity comparison.

If both pairs have the same separation, i.e.~$x_1 + \delta$ and  $x_2 + \delta$, with $x_2>x_1$, it follows that:
\begin{align}
    \mathcal{J}(x_1,x_1+\delta) = \frac{x_1}{x_1+ \delta }  < \frac{x_2}{x_2+ \delta} =  \mathcal{J}(x_2,x_2+\delta)
\end{align}
This result means that the Jaccard similarity comparison is more strict when performed between pairs of equally separated values that are closer to the origin of the coordinate system. 

As a means to complement the quantification of the similarity comparison, the \emph{coincidence similarity index} has been suggested, which, in the case of one-dimensional features, can be defined as:
\begin{align}
  \mathcal {C}(x,y) = \left[  \mathcal{J}(x,y) \right]^D 
\end{align}
where $D$ is an exponent that controls how strict the similarity comparison is performed. The higher the value of $D$, the more strict the comparison becomes.

The Jaccard and coincidence indices are of special interest when the measured data has a proportional structure.  When applied in these situations, the comparison between proportionally separated pairs of values will yield the same result.

Figure~\ref{fig:prox_sim}(b) exemplifies the similarity values obtained when comparing $x$ with $x_r=1$, $x_{min}=0$ and $x_{max}=5$. Interesting properties can be observed, including: (i) all values take the maximum value 1 at $x = x_r$; (ii) the curves are not symmetric around $x_r$; (iii) the implemented comparisons are more strict for $x<x_r$; (iv) the coincidence index is more strict than the Jaccard index; and (v) larger values of $D$ imply more strict comparisons.

\section{A Case Example}

This section presents a case example intended to help the understanding of the concepts discussed in the present work.  It refers to situations in which the type of measurement, transformations, normalizations, and comparisons need to be chosen in order to satisfy specific research or application situations, specifically the relationship between the variation and the magnitude of the quantity.  

The following specific example concerns a hypothetical situation in which the weights of several species of fruits, represented in terms of the random variable $X$, are to be compared within each respective group.  In a preliminary step, samples of fruits were measured linearly in grams, and the average and standard deviation of each values obtained for each species of fruit were obtained. In the case of this example, the choice of grams was made from the outset because the traditional approach which is typically in similar circumstances, also taking into account that the available instruments (scales) commonly provide linear quantifications of weight in grams.

Since the measurement procedure and unit were fixed from the beginning, what remains is to devise a particularly interesting approach to compare the obtained measurements.

\begin{figure}[ht]
  \centering
     \includegraphics[width=0.55 \textwidth]{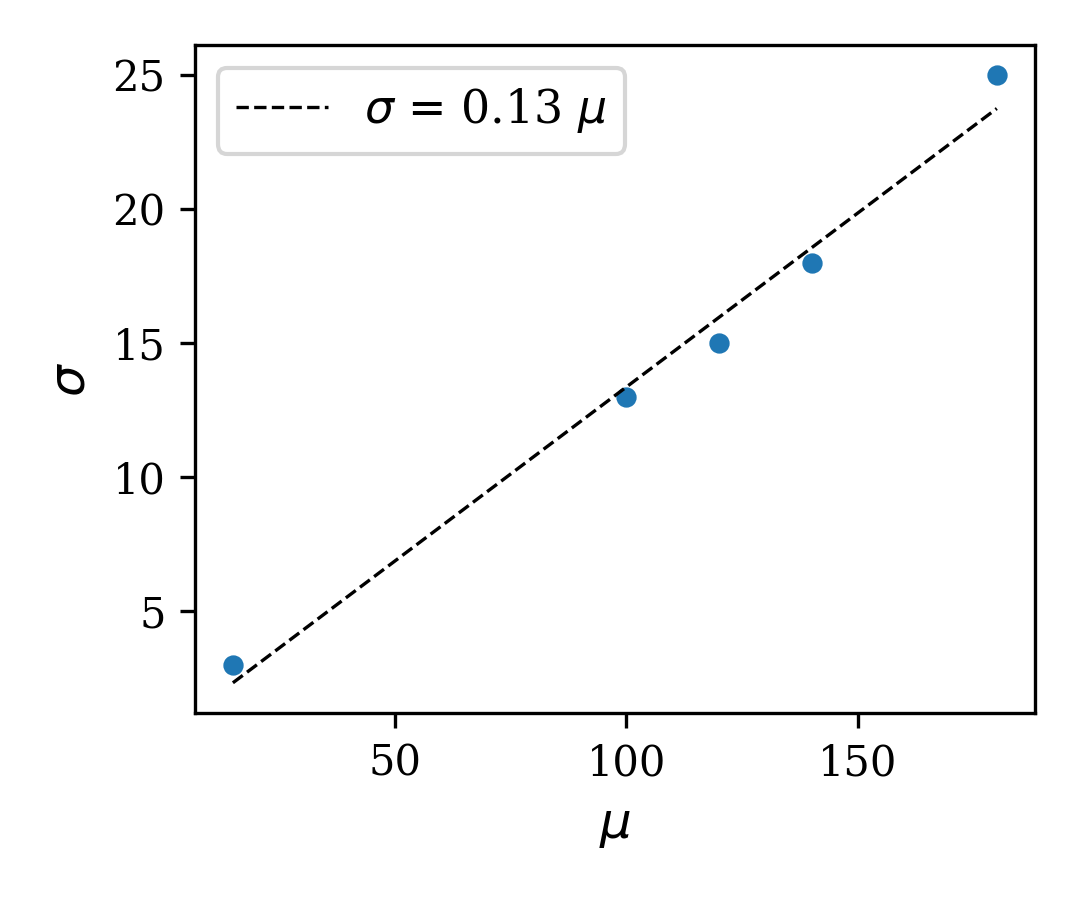}
  \caption{The relationship between standard deviation and average of five species of fruits considered in the case example describe in the present section.}
  \label{fig:fruit}
\end{figure}

When the standard deviation was plotted in terms of the average for each species, they were found to present the relationship shown in Figure~\ref{fig:fruit}, including the relationship $\sigma =0.13 \mu$ obtained from the respective linear regression.

It can be seen from the plot that a nearly linear relationship exists between the standard deviation and average of weights for the considered species of fruit.  This suggests a \emph{proportional relationship} between the variability and the average of the measured quantity.  

\begin{figure}[ht]
  \centering
     \includegraphics[width=0.55 \textwidth]{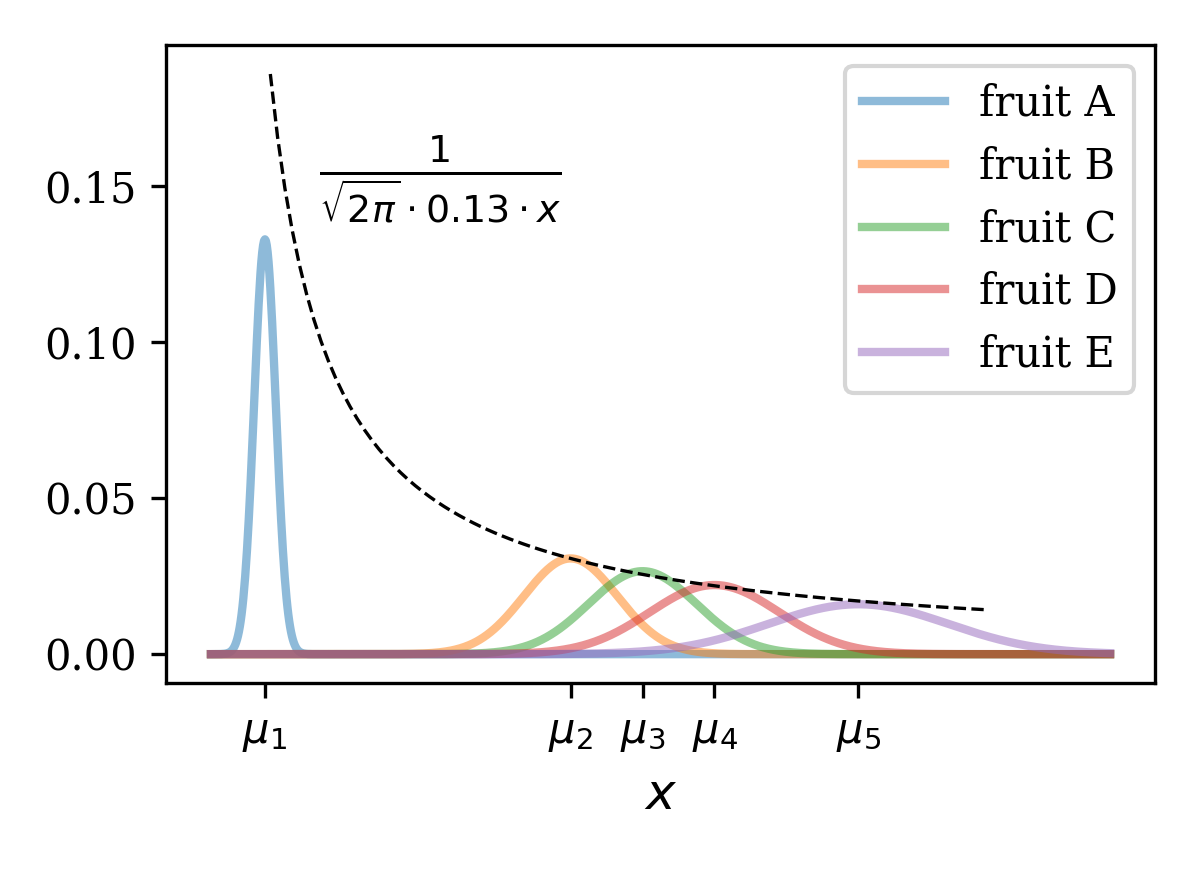}
  \caption{Illustration of the normal densities respective to each of the fruit species in Fig.~\ref{fig:fruit}. The assumed relationship between the standard deviation and the average is also shown in terms of the dashed line.}
  \label{fig:fruit2}
\end{figure}

Figure~\ref{fig:fruit2} illustrates the normal densities corresponding to the fruit species in Figure~\ref{fig:fruit}.

It is interesting to observe that the proportionality between the dispersion and average magnitude provides a means for the respective normal densities involving negligible tail extending along the negative portion of the horizontal axis.  

There is another important aspect to be considered, namely that the statistical model (e.g.~\cite{johnson:2002,Bertsekas}) observed by the fruit weight measurement may not be completely modeled in terms of standard deviation and average, which is the case of non-normal densities. In those cases, the observed proportional relationship can only be considered in an approximate sense.  In the following, it is assumed that the densities describing the weight of each species of fruit are well-approximated by normals.

The observed proportionality between the standard deviation and average suggests that, if the weights of fruits of a same species are to be compared among themselves in an absolute manner, a larger variation of obtained values would be obtained in the case of heavier fruits than otherwise.  Indeed, this approach is supported by the verification that the standard deviation closely scales proportionally with the average magnitude of the measurement (s shown in Fig.~\ref{fig:fruit}).  

Now, if one wants to relate the comparisons obtained for each species of fruit, it could be interesting to do so in a way that the variation became constant with the magnitude of the measurements.  This would allow, for instance, the identification of common values of threshold for grading the fruits, in the sense that a specific sample of fruit would be understood as being particularly large if its weight is larger than a given proportion common to all species.  

Going back to our hypothetical example, one could consider at least the two following possibilities for how to proportionally compare the weights within each of the species.  In order to provide a specific example, we consider the comparison between the weights of pairs of fruits of species A and E, namely $w_{A,1}$ and $w_{A,2} =  2 \, w_{A,1}$ and $w_{E,1}$ and $w_{E,2}= 2 w_{E,1}$.  We would expect both comparisons to yield the same value.  The two possibilities are:

I:  Transform the measured values by taking their logarithm.  This can be done by taking the measurements in grams and then calculate the logarithm, or by using a scale which already yields the logarithm of the weights.  Therefore, the measurement unit would be changed. However, this would allow the values to be compared by using a linear approach such as the Euclidean distance between weight values.  The disadvantage of this choice is that the comparison would take on logarithms of the original measurements, while humans are familiar with considering weights in grams.  In the case of the example, we would have:  $\log(w_{A,2})-\log(w_{A,1}) = \log(w_{E,2})-\log(w_{E,1}) = \log(2)$.

II: Keep the measurements in grams, without additional transformations or normalizations, but use a proportional approach, such as the Jaccard similarity index, to compare the values.  This approach has the advantage of keeping the original measurement unit and linearity up to the moment of comparison, which in this case would be dimensionless and normalized in the range $[0,1]$.  Unlike the previous cases, the input to the comparison would be the weights in grams, and not a transformed quantity.  Observe that no transformation or normalization is required.  In addition, the Jaccard similarity has an intuitive explanation in terms of set theory, more specifically as the ratio between the intersection and union. In the case of comparing values in continuous spaces as the $\Re$, the set operations of intersection and union translate, respectively, as the minimum and maximum between the two compared values.  This means that the weights can be compared while in their original units of grams, with a intuitive dimensionless result being obtained.  In the case of the considered specific examples, we have $\mathcal{J}(w_{A,2},w_{A,1}) = \mathcal{J}(w_{E,2},w_{E,1}) = 1/2$.

Although the choice among these possibilities would ultimately depend on each specific situation, the described example nonetheless illustrates the several interrelationships between the constraints and concepts involved.

\section{Concluding Remarks}

Both science and technology are founded on the quantification, through respective measurements, of properties of real-world or abstract quantities. Then, once quantities are measured, they often need to be related or compared. In a given situation, the choice of the measurement and comparison approach to adopt may not be particularly straightforward. Some of the many aspects underlying the measurements and their comparison have been presented and discussed in this work.

After preliminary considerations about the properties of measurements, three main categories have been presented and discussed respectively to the measurement variation or interval of interest $\delta$: uniform, proportional, and inverse measurements. After discussing some of their main properties and possible normalization, three types of comparisons have been presented and discussed, namely Euclidean distance, proximity (three possible approaches), and similarity (Jaccard and coincidence).

In addition to several more specific identified properties presented by each of the considered types of measurement and comparisons, one more general conclusion is that the values of measurements, normalization, and comparisons can vary significantly with the choice of the methods and respective parameters. That is why special attention must be paid while choosing specific approaches, which should take into account, among other aspects, the statistical distribution of the observed values, how the measurement variations and scales of interest tend to vary with the respective magnitude, the available means for implementing the measurements (e.g.~instruments or procedures), as well as specific questions and requirements implied by each experiment and respective study.  A case example concerning the comparison of fruits of same species was also included in order to illustrate the interplay between the several choices and requirements that can be involved in such situations.

Given that the present work focused on one-dimensional values, it would be interesting to complement this study by considering higher dimensional data, which imply additional aspects to be taken into account. In addition, it would be interesting to consider other types of measurement, normalization, and comparison.

\section*{Acknowledgments}
Luciano da F. Costa thanks CNPq (grant no.~313505/2023-3) and FAPESP (grant 2022/15304-4). A. Benatti is grateful to FAPESP (grant 2025/26083-7 and 2022/15304-4).

\bibliographystyle{unsrt}
\bibliography{mybib}

@book{Bertsekas,
  author      = {D. P. Bertsekas and J. N. Tsitsiklis},
  title       = "Introduction to Probability",
  publisher   = "Athena Scientific",
  year        = 2008
}

@book{johnson:2002,
author = "Johnson, R. A. and Wichern, D.W.",
title = "Applied multivariate  analysis",
publisher = "Prentice Hall",
year = 2002
}

@Article{Blizard,
  Title                    = {Multiset Theory },
  Author                   = {W. D. Blizard},
  Journal                  = {Notre Dame Journal of Formal Logic},
  Year                     = {1989},
  Pages                    = {36---66},
  Volume                   = {30}
}

@Article{Blizard2,
  Title                    = {The Development of Multiset Theory},
  Author                   = {W. D. Blizard},
  Journal                  = {Modern Logic},
  Year                     = {1991},
  Pages                    = {319--352},
  Volume                   = {4}
}

@article{benatti2024,
  title={Normalization in proportional feature spaces},
  author={Benatti, A. and da F. Costa, L.},
  journal={arXiv preprint arXiv:2409.11389},
  year={2024}
}

@article{benatti2024partially,
  title={Partially Proportional and Adaptive Similarity Indices},
  author={Benatti, A. and da F. Costa, L.},
  journal={arXiv preprint arXiv:2410.17894},
  year={2024}
}

@misc{Jac:wiki,
  author = {Wikipedia},
  title = "Jaccard Index",
  year={2021},
  note= {\url{https://en.wikipedia.org/wiki/Jaccard_index}.
[Online; accessed 24-Sep-2026].}
}

@misc{da2021further,
  title={Further Generalizations of the {J}accard Index},
  author={da F. Costa, L.},
  year={2021},
  journal = "ResearchGate",
  howpublished= "\url{https://www.researchgate.net/publication/355381945_Further_Generalizations_of_the_Jaccard_Index}",
}

@book{snyder1991,
  title     = {Random Point Processes in Time and Space},
  author    = {Snyder, D. L. and Miller, M. I.},
  year      = {1991},
  publisher = {Springer-Verlag},
  address   = {New York}
}

@book{Chertov1964,
  title     = {Units of Measurement of Physical Quantities},
  author    = {Chertov, A. V.},
  year      = {1964},
  publisher = {Hayden Book Company},
  address   = {New York}
}

@book{morris2020,
  title={Measurement and Instrumentation: Theory and Application},
  author={Morris, A. S. and Langari, R. and others},
  year={2020},
  publisher={Elsevier},
  address   = {San Diego, CA}
}

@book{cataldo2020,
  title     = {Basic Theory and Laboratory Experiments in Measurement and Instrumentation: A Practice-Oriented Guide},
  author    = {Cataldo, A. and Giaquinto, N. and De Benedetto, E. and Masciullo, A. and Cannazza, G. and Lorenzo, I. and Nicolazzo, J. and Meo, M. T. and De Monte, A. and Parisi, G. and Gaetani, F.},
  year      = {2020},
  publisher = {Springer Nature Switzerland},
  address   = {Cham, Switzerland}
}

@book{baird1995,
  title     = {Experimentation: An Introduction to Measurement Theory and Experiment Design},
  author    = {Baird, D. C.},
  year      = {1995},
  edition   = {2nd},
  publisher = {Prentice Hall},
  address   = {Englewood Cliffs, NJ}
}

@book{holman2012,
  title     = {Experimental Methods for Engineers},
  author    = {Holman, J. P.},
  year      = {2012},
  edition   = {8th},
  publisher = {McGraw-Hill},
  address   = {Boston, MA}
}

\end{document}